# Rayed craters on Dione: Implication for the dominant surface alteration process

Authors

**Naoyuki Hirata[a,*], Hideaki Miyamoto[b,c]**

[a] Graduate School of Science, Kobe University, Rokkodai, Kobe 657-8501, Japan.
[b] The University Museum, The University of Tokyo, Hongo, Tokyo 113-0033, Japan.
[c] Planetary Science Institute, Tucson, AZ 85719, USA
[*] Corresponding Author Email address: hirata@tiger.kobe-u.ac.jp

**Proposed Running Head:** Rayed craters on Dione

**Editorial Correspondence to:**
Dr. Naoyuki Hirata
Kobe University
Rokkodai 1-1 657-0013
Tel/Fax +81-7-8803-6685
Email: **hirata@tiger.kobe-u.ac.jp**

Highlights
► We examine the patterns and spatial distributions of rayed craters on Dione.
► We find that the ray system of Creusa crater extended over Dione.
► The rays deposited on the trailing hemisphere have been partially erased.
► This is due to implantation of dark particles, found throughout Saturn system.
► Our observations imply that the implantation event occurred very recently.

**Abstract**

From recently-acquired, high-resolution images obtained by the Cassini spacecraft, we examine the patterns and spatial distributions of rayed craters on Dione. We identify 29 rayed craters with diameters larger than 2km on Dione's surface. The density of rayed craters and theoretical cratering rates indicate that the retention time for rays on Dione can be approximately 1-50 My. Such a short retention time is interpreted to be due to bombardment of plasma and E-ring particles, as well as implantation of dark particles (presumably the same dark material found on Hyperion, Iapetus, and other saturnian satellites). We also find that when the ray system of Creusa crater was formed, it extended over most of the surface of Dione. Later, the ray system deposited on the trailing hemisphere might have been partially erased, mostly due to implantation of dark particles, which may have also removed other bright ray systems in that region. The pattern of Creusa's ray system implies that the implantation of the dark material occurred more recent than both the age of Creusa crater and the typical retention time for rays on Dione.

## 1. Introduction

Dione is a mid-sized satellite of Saturn with a radius of 561 km. It orbits Saturn every 65.7 hours in synchronous rotation at a distance of 377,396 km. One of the most notable features on Dione are ray systems around numerous craters (Wagner et al. 2011), as shown in Fig. 1. Rays are generally bright deposits ejected from a newly-formed impact crater that gradually fade with time due to various surface alterations (e.g. Melosh 1989, Shoemaker 1962). The most extensive ray system on Dione, which is centered about the largest rayed crater, 36.2-km-diameter Creusa (located at 76°W, 49°N; Fig. 1a), extends for at least several hundred kilometers across most of the leading hemisphere (Wagner et al. 2011); it has been identified as one of the youngest geological features on Dione (Stephan et al. 2010, Scipioni et al. 2013).

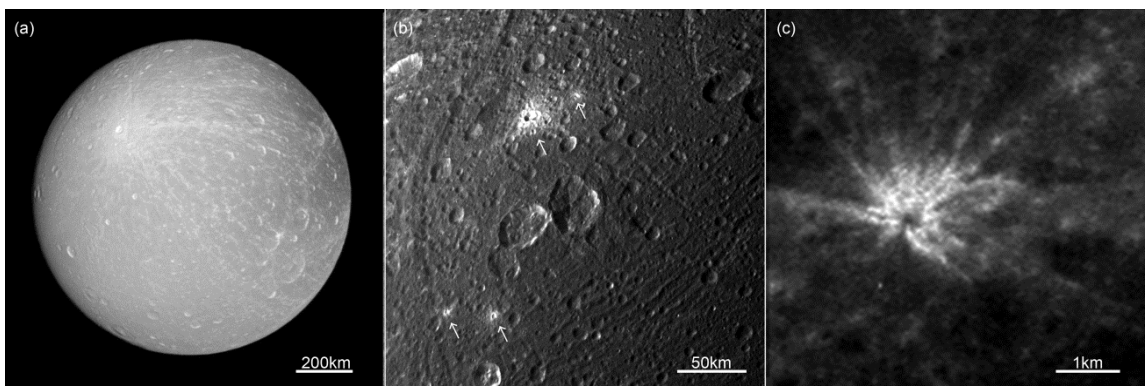

**Fig. 1.** Rayed craters on Dione. (a) The ray system of Creusa crater (PIA14628), which is known as the largest rayed crater on Dione. (b) At least 4 rayed craters (white arrows) can be seen on this image (N1711606439). The center of this image is 25.7°N, 128.8°W. (c). A small rayed crater (300 meters in diameter; at 36.8°S, 198.3°W; N1507745801). This is the smallest rayed crater which we can identify.

The distribution and the density of rayed craters can be good indicators of both the current cratering rate and the surface-alteration rate. The former depends on the fluxes of (i) trojan objects of the gas giants, (ii) Centaurs and ecliptic comets, (iii) nearly isotropic comets, (iv) potentially a few main-belt asteroids, (v) Saturn's irregular satellites, and (vi) secondary or sesquinary impactors (Dones et al. 2009). On the other hand, the latter is a result of the combined mechanisms of (1) charged particle bombardment, (2) E-ring particle accumulation, (3) implantation of sub-micron dark particles, (4) micrometeoroid bombardment, (5) UV or cosmic ray irradiation, and (6) sublimation (Schenk et al. 2011, Jaumann et al. 2009, Clark et al. 2008).

### 1.1. The distribution of rayed craters

Determining the distribution of rayed craters is useful for constraining the dominant impactor populations and primary surface-alteration processes. This is because the relative rates of impactors or surface alterations vary depending on latitude or longitude. Heliocentric impactors (i~iv) preferentially impact on the leading hemisphere, while planetocentric impactors (v and vi) do not cause such an asymmetry (Horedt and Neukum 1984, Zahnle et al. 2001). Charged particles, such as plasma ions and electrons (<1 MeV), primarily affect the trailing hemisphere because Saturn's magnetosphere co-rotates at a rate faster than the orbital speed of the moons. Conversely, electrons above roughly 1 MeV in energy preferentially bombard the equatorial region of the leading hemisphere (Paranicas et al. 2014). E-ring particles, generated by the plume from Enceladus, preferentially accumulate on the leading hemisphere (Ostro et al. 2010, Verbiscer et al. 2007). The trailing hemisphere of Dione appears to be a dark surface formed by the bombardment of sub-0.5-μm diameter dark particles impacting the satellite from the trailing-side direction (Clark et al. 2008, Stephan et al. 2010). The dark particles could be the dark material (referred as the low albedo material) found widespread throughout the Saturn system (such as observed on both Hyperion and Iapetus (Clark et al. 2008, Stephan et al. 2010)). UV irradiation and sublimation preferentially affect the surface at relatively low latitudes because these phenomena depend on the solar incident angle. On the other hand, cosmic-ray irradiation symmetrically affects the surface.

Schenk and Murphy (2011) mapped the global distribution of rayed craters using the color-ratio maps. They showed that the density of rayed craters is significantly higher (by a factor of ~4) on Dione's leading hemisphere compared to its trailing hemisphere. We note that images at high sun (i.e., solar incident angles are close to normal to the surface) are also suitable to identify the distribution of rays or rayed craters because they can significantly enhance brightness contrasts (Fig. 2). An example of this elsewhere in the solar system is the ray system of Tycho on the Moon, which is obvious at full moon while hardly visible under lower illumination. Therefore, we generate a new map of high sun images to examine the distribution of rayed craters and the ray system of Creusa, which may reflect local relative rates of the retention time for rays.

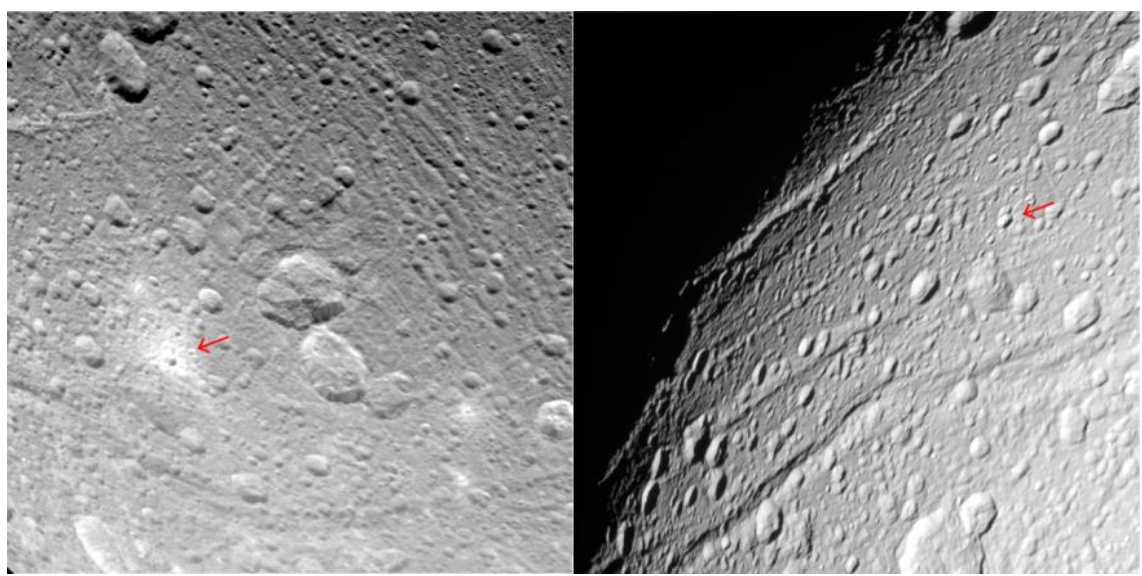

**Fig. 2.** A rayed crater of Dione taken at high sun (left: N1711606439) and low sun (right: N1507741140). Arrows indicate the same rayed crater. Solar incident angles of these images are, respectively, 25.5 and 76.1 degrees. The visibility of ray system is strongly dependent on solar incident angle.

**1.2. The density of rayed craters**

The density of rayed craters helps to assess the rate of major impactors and the extent of their alteration of the planetary surface. Specifically, the retention time for rays, suggested by the fraction of rayed craters out of the total impact crater population, can be a good indicator of the geologic timescale of surface alteration. Examples of this include the fractions of rayed craters on the Moon and Mercury, which are roughly a tenth of the total impact populations (Allen 1977), interpreted to indicate that ray systems of terrestrial planetary bodies fade out during at most 1 Gy. Similarly, the retention times for rayed craters on Ganymede and Callisto are roughly estimated to be 1-2 Gy, based on the global distribution of rayed craters (Passey and Shoemaker 1982).

The rayed craters on Rhea and Dione are estimated to be much younger than 500 Ma (Wagner et al. 2011). This estimate is based on the crater size-frequency distribution of the floor of Inktomi crater, the largest rayed crater on Rhea (Wagner et al. 2008). However, other rayed craters on Dione and Rhea are not imaged at sufficient resolution for crater counting. In addition, the crater size-frequency distribution of Inktomi crater cannot by itself constrain the retention time for rays. Instead, we examine the fraction of rayed craters out of the total crater population of Dione, which can provide better constraints for the retention time for rays and the age of Creusa crater.

2. Method

As of December 2013, 1881 images (higher than 5 km/pixel) of Dione have been obtained by the ISS camera onboard the Cassini spacecraft and are available via NASA's Planetary Data System. The entire surface of Dione has been imaged at a resolution of 700 m/pixel or better.

2.1. A new map of high sun images

Global maps of Dione have been developed by other researchers (e.g. Schenk et al. 2011). They are most often generated from images with higher solar incident angles, which are most useful to identify topographic features. However, these are not best suited for our purpose because a brightness contrast is generally not identifiable from those images. Therefore, we generate a new map by mosaicking images whose solar incident angles are close to normal to the surface (Fig 3e).

We use ISIS3 software produced by the U.S. Geological Survey (USGS) for projecting images, radiometric calibrations, and crater measurements. We describe the image processing to develop the mosaic in Supplementary material. Our newly developed cylindrical map (Fig. 3e) has enhanced brightness contrasts on the surface of Dione, and thus is useful for rough identifications of brightness differences (even though imaging conditions vary slightly at locations, and thus do not reflect the exact values of albedo). Using this global mosaic, we map the distribution of rays on Dione including potential rays from Creusa crater (Fig. 3f). Note that we use raw images for clarifications, particularly for distinguishing exposures of linear cliffs from rays.

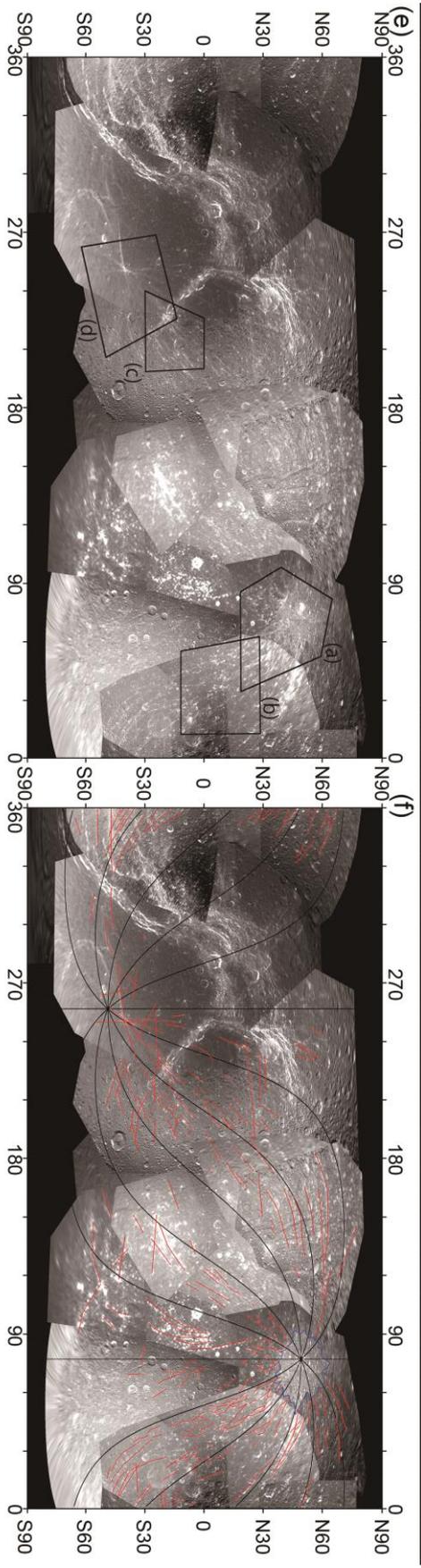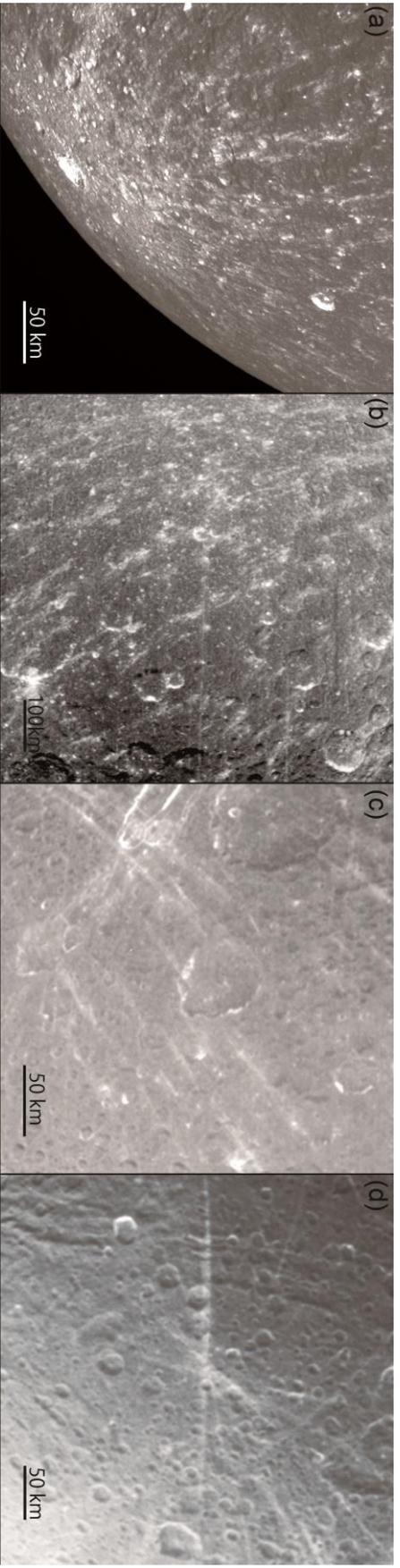

**Fig. 3.** (a) Creusa crater and its ray system (N1649319166). (b) Bright linear structures on the leading hemisphere (N1649318703). (c) Bright linear structures on the trailing hemisphere (N1643300999). (d) Bright linear structures on the antipode of the Creusa crater (N1501620702). (e) Locations of images on the cylindrical projection map using the images where the solar incident angle is close to perpendicular. The following 17 images are used: N1711614785, N1696243718, N1643298118, N1569837386, N1569828482, N1569828604, N1563783994, N1532405021, N1569854105, N1593074283, N1514077920, N1649317673, N1643300999, N1581274186, N1581192726, N1649318982, and N1649331848. (f) The distribution of bright linear structures (red curves) on the cylindrical projection map. The region surrounded by a blue line represents Creusa crater and its ejecta. The black lines highlight the orthodromic distance between Creusa crater and its antipodal point. The distribution shows that bright linear structures are the ray system from Creusa.

### 2.2. Measurements of rayed craters

Rays at high latitudes (about 45° or higher) are difficult to study because the orbit of Dione always results in a large solar incident angle in these regions. In addition, rayed craters in the vicinity of Creusa crater (about 20° in the angular radius centered at Creusa) are also difficult to identify because of the bright ejecta from the Creusa impact event. Thus, overall, we find that rayed craters are detectable on about 69 % of the total surface of Dione (hereafter, the detectable region; Fig. 4a).

Craters have circular raised rim and an interior slope that is steepest near the rim and smoothly decreases toward the crater's center, which is surrounded by an ejecta deposit (Melosh 1989). Rays are bright albedo features that extend from fresh craters in subradially to radially orientated filaments (Melosh 1989). In this work, we identify rayed craters based on a circular depression and bright albedo features. We use images both at high sun and low sun for each crater because we can judge bright albedo features based on images at high sun, and topographic features such as depression and rim based on images at low sun (Fig. 1 and 2). We note that a middle sun image often allows us to identify both ray and depression, and then we judged ray craters in the single image.

To measure the diameter of craters, we use an application "qview" in ISIS3, which can display images for interactive analysis and measure the distance between any two points in the images. We consider that almost all rayed craters (> 2km in diameter) can be identified in the detectable region because the resolutions of images are 400 m/pixel or better. These resolutions also imply that diameters of craters in this work would be in error by less than 400 meter. In the case of a non-circular crater, we define that the major axis of the crater represents its diameter. The distribution of identified rayed craters is shown in Fig. 4. Fig. 5 shows the cumulative size-frequency distribution of the identified rayed craters on Dione. Error bars are defined as $\pm R*N^{-0.5}$, where N is the number of craters in a bin. For these analyses, we follow the method of the Crater Analysis Technique Working Group (1979).

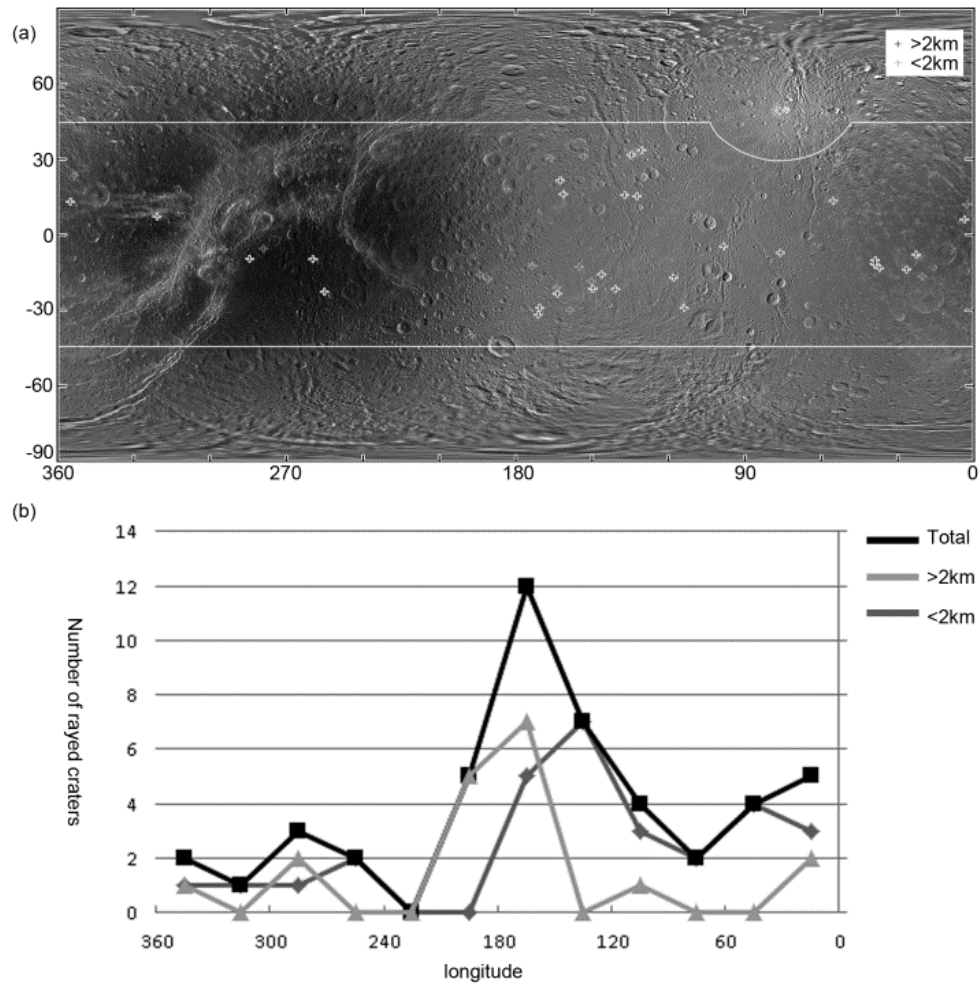

**Fig. 4.** (a) The distribution of rayed craters on Dione. Black and gray dots represent, respectively, rayed craters larger than 2 km and ranging between 1.5 km and 2km in diameter. White lines outline the detectable region for rays (see text). We use PIA18434 for a background projection map. (b) The longitudal dispersion of rayed craters.

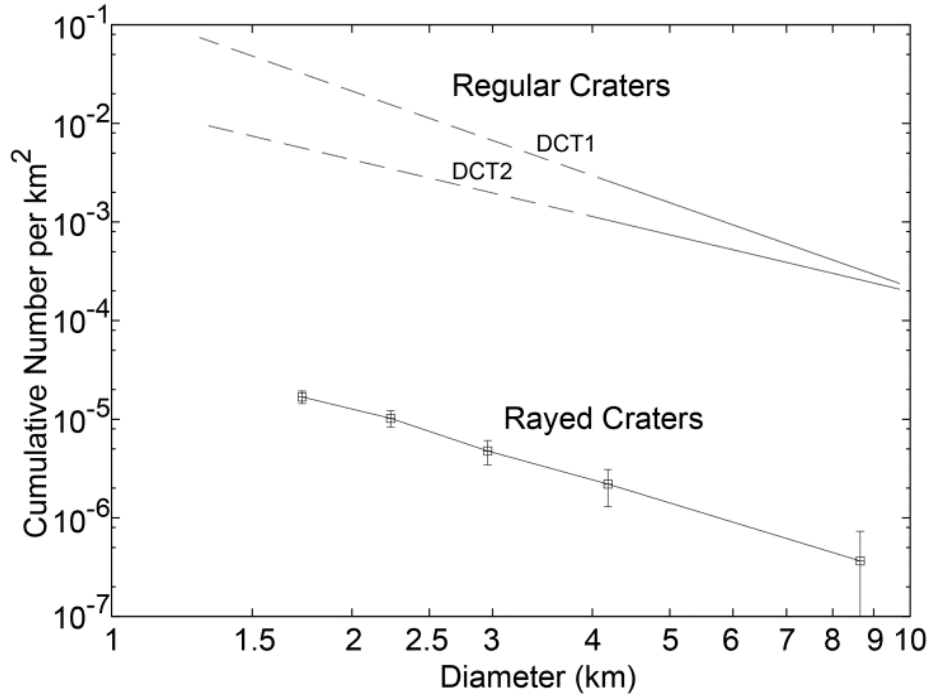

**Fig. 5.** The cumulative size-frequency distribution of regular craters and rayed craters on Dione. The data of regular craters comes from Kirchoff and Schenk (2015). DCT1 and DCT2 represent the crater density of Dense Cratered Terrain 1 and 2 defined by Kirchoff and Schenk. Kirchoff and Schenk database deals with regular craters larger than 4 km, and therefore, D < 4km (the dashed lines) are extrapolated from the database.

## 3. Result
### 3.1. The ray system of Creusa crater

The global distribution of large, bright linear features (Fig. 3f) implies that these are rays from Creusa crater and that the ray system extends from the crater's margin to its antipode, except for within the trailing hemisphere (especially at the trailing apex). This suggests that the ray system from the crater, and stretched over most of Dione at the time of its formation, with subsequent destruction of rays in the trailing hemisphere. We note that a ray system that extends a full 360° across a planetary surface has not yet been identified on any other solar system body. We also note that other origins have been suggested for the bright linear structures on the antipode of the Creusa crater (Fig. 3d). Plescia (1983) interpreted them as a small rayed crater called Cassandra crater in a Voyager-era study, while Wagner et al. (2006) argued they may be a set of radial scarps formed by tectonism from Cassini images. However, we can identify neither a small rayed crater nor a set of scarps (Fig. S1 in Supplementary material).

## 3.2. Other small rayed craters

We find 29 rayed craters with diameters larger than 2km (D>2km) on Dione, as shown in Fig. 4a and Table S1 in Supplementary material. We note that at least 47 rayed craters are larger than 1.5 km, and some of these may be larger than 2 km because of the measurement error. Extrapolating the Kirchoff and Schenk (2015) crater measurements for D > 4km over 85% of Dione's surface, the dense cratered terrains on Dione for a D>2km population is estimated to be 4,000 ~ 20,000 craters/$10^6 km^2$. On the other hand, we find the crater density of D>2km rayed craters on Dione is 10.3 craters/$10^6 km^2$, assuming 28 rayed craters per $2.73 \times 10^6$ $km^2$ (equal to 69% of Dione's surface). This indicates that the fraction of rayed craters out of the total crater population is less than 1%. The density of rayed craters on the trailing hemisphere is one-fifth as dense as that on leading hemisphere, which is consistent with the result of Schenk and Murphy (2011). Most of rayed craters favor a prime meridian or the 180th meridian, rather than the leading hemisphere (Fig. 4b).

## 4. Discussion
## 4.1. The retention time for rays

Rayed craters with D>2km do not appear to be related to secondary impacts from Creusa or sesquinary impacts from Inktomi crater. This is because the maximum size of secondary craters is at most 4% of the primary diameter on the Moon, Mars, and elsewhere (Melosh 2011). Most of small rayed craters (D < 1.5 km) may have originated from the Creusa impact event.

Cratering rates are given for primary craters with a heliocentric origin, but not for secondary or sesquinary craters. The heliocentric cratering rate in the Saturn system is theoretically estimated by Zahnle et al. (2003) and is based upon various observational data in the outer Solar system and assuming comets are the dominant impactors. Moreover, Zahnle et al. (2003) considers two cases for Saturn: Case A (based on comet sizes at Jupiter) and Case B (based on comet sizes at Triton).

The theoretical rate on Dione for D>2 km (Zahnle et al. 2003) is $2.3 \times 10^{-13}$ crater/$km^2$·year (Case A) or $6.3 \times 10^{-12}$ crater/$km^2$·year (Case B). Thus, the density of rayed craters indicates a retention time of about 44.6 or 1.63 My, respectively. Similarly, we can obtain a retention time on the leading hemisphere of 75.2 My or 2.74 My (assuming 23 rayed craters per $1.33 \times 10^6$ $km^2$) and that on the trailing hemisphere of

15.5 My or 0.57 My (assuming 5 rayed craters per 1.40x10$^6$ km$^2$). In summary, the typical retention time for rays on Dione is estimated to be approximately 1-50 My, which is shorter than the estimate suggested by Wagner et al. (2011).

Based on the above estimates, the retention time for rays on the leading hemisphere is longer than that on the trailing hemisphere. However, the production rate of heliocentric impactors at the apex of motion (the center of the leading hemisphere) is estimated to be 15 to 100 times greater than that at the antapex (the center of the trailing hemisphere) (Horedt and Neukum 1984, Zahnle et al. 2001), even though strong asymmetries have not been identified on any of the satellites (Zahnle et al. 2001, Dones et al. 2009). If we assume the above differential cratering rate, no major difference exists in the retention time between both hemispheres.

**4.2. The hemispheric asymmetry of rayed craters**

Schenk and Murphy (2011) suggested that the asymmetric distribution of rayed craters is potentially explained by secondary impacts from Creusa or sesquinary impacts from Inktomi, heliocentric impactors, or the relative rates of surface alteration on the trailing and leading hemispheres. We consider that neither secondary nor sesquinary craters are responsible for the asymmetric distribution because of their sizes as discussed above. Heliocentric impactors cannot be ruled out, as we discuss above. Yet, this cannot likely be the only reason for the asymmetry because the lack of the ray system of Creusa on the trailing hemisphere cannot be explained. Instead, the asymmetry can be best explained by the differential resurfacing rate, which is evidenced by the lack of the ray system of Creusa on the trailing hemisphere.

Perhaps, the retention time for rays in the leading hemisphere is largely different from that in the trailing hemisphere. If so, what kinds of surface alterations can cause the asymmetric distribution of rayed craters? Micrometeoroid bombardment and UV or cosmic ray irradiation cannot be the only processes of ray erasure because these phenomena do not preferentially affect the surface of the trailing hemisphere. Sublimation may contribute to the shorter retention time on the trailing hemisphere because the trailing hemisphere has lower albedo. In fact, craters on slowly rotating satellites with low-albedo surface, such as Iapetus, Ganymede, and Callisto, are predominantly and quickly eroded due to temperature-driven, water-ice sublimation (Denk et al. 2010, Prockter et al. 1998, Squyres 1980). However, this effect on Dione must be minor because, according to Spencer and Denk (2010), sublimation rates are negligible for the retention time for rays on Dione.

Rather than these mechanisms, we consider that implantations of dark particles,

E-ring grain accumulations, and magnetospheric plasma bombardments are plausible. These mechanisms are known to primarily lead to Dione's global color pattern at the present time (Schenk et al. 2011). Especially, the enhancement is significantly stronger on the trailing hemisphere where it is associated with the dark material (Schenk et al. 2011). On the other hand, the prime meridian and the 180th meridian of Dione are hardly affected by them. This is in good agreement with the longitudal distribution of rayed craters (Fig. 4b). This view is also supported by the ray system of Creusa, which is likely to become faded toward the trailing apex where there appears to be a higher abundance of dark particles (Fig. 6).

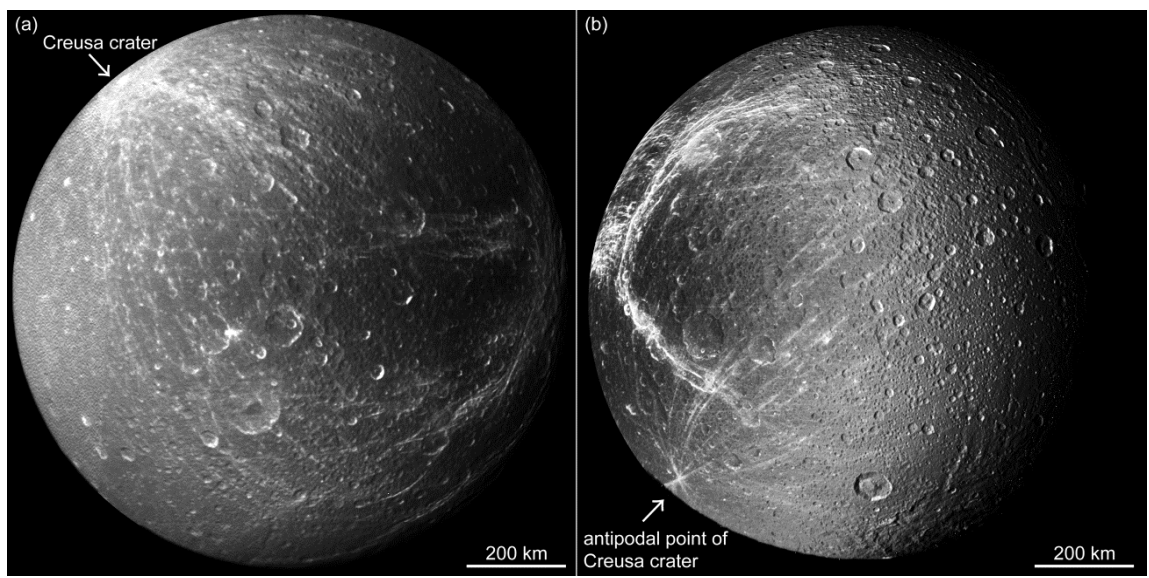

**Fig. 6.** (a) Near side of Dione. The center of this image is at 0°N, 350°W. Creusa crater can be seen in upper left. (b) Far side of Dione. The center of this image is at 0°N, 170°W. This image is the exact opposite side of (a). The antipodal point of Creusa crater is in the lower right of the image. The following 6 images are used: W1649313601, W1649316019, N1643297347, N1643298118, N1643300578, and N1643300999.

5. **Conclusions**

Overall, the retention time for rays on Dione is as long as several tens My. The maturation of materials on Dione seems to occur very fast, which reflects the timescale of surficial chemical or physical alterations presumably due to E-ring particle, sub-micron dark particle, and/or magnetospheric plasma bombardments. We find the three processes control not only the global color pattern of Dione but also the spatial distribution of rayed craters. It is unclear whether implantation of the dark particles is a continuing phenomenon or a transient past event. In any case, the above results imply that this process would have occurred relatively recently compared to the retention time for rays or the formation time of Creusa crater.

**Supplementary material**

This article contains additional supplementary material: Table S1, Figure S1 and S2, and description for the image processing.

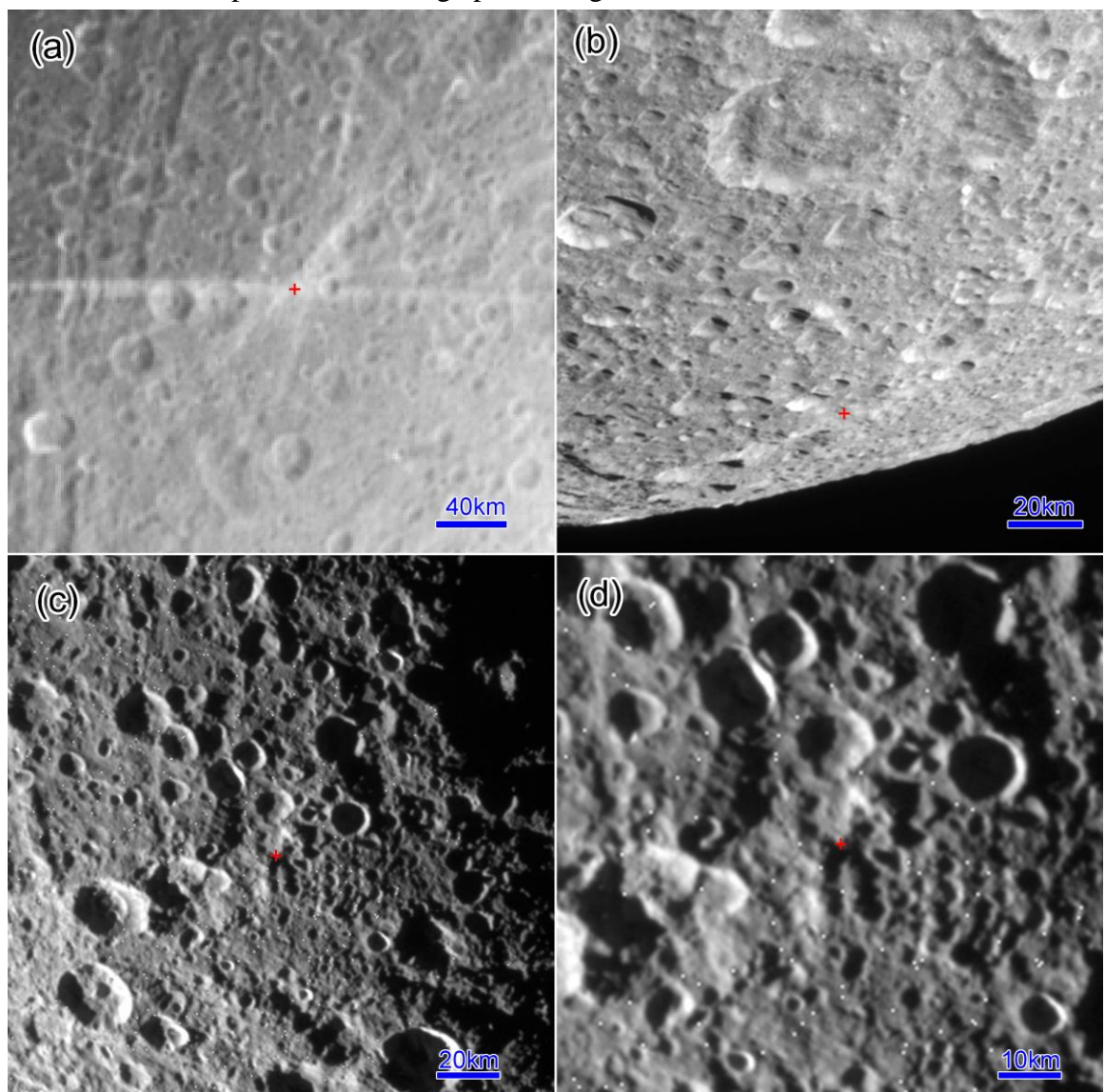

**Figure S1.** The antipodal points of Creusa crater at high sun (a), oblique view (b), and low sun (c, d). Red points shown in (a) to (d) indicate the same place. High sun image (a) shows bright linear structures with 5~10km in width, whereas low sun images (c, d) and oblique image (b) appear no landform associated with slope. Here we use N1501620702 (1.4km/pixel) for (a), N1662202200 (284m/pixel) for (b), and N1569814968 (410m/pixel) for (c, d).

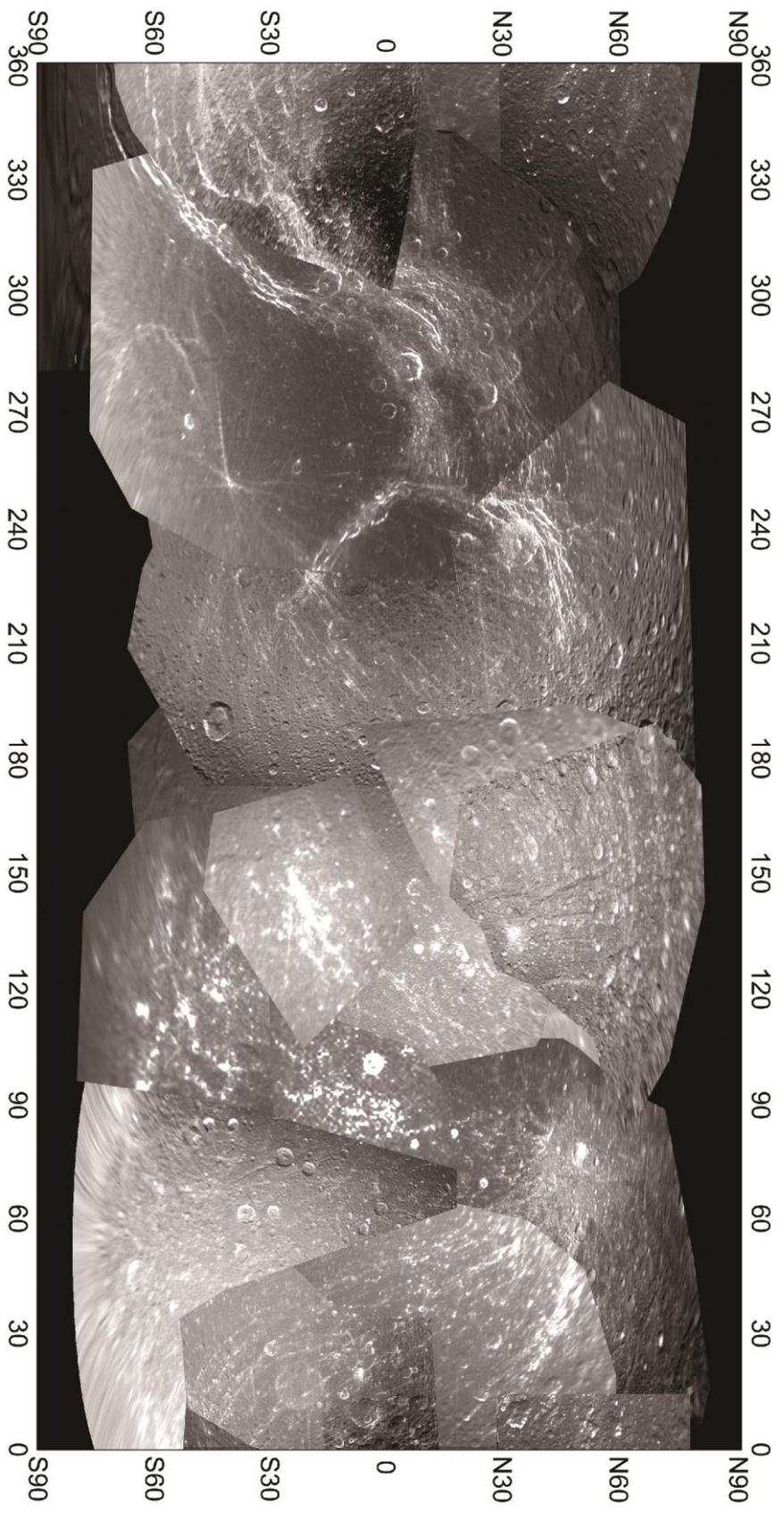

**Figure S2.** The simple cylindrical projection map using the images where the solar incident angle is close to perpendicular.

**Supplementary Description: Procedure of the image processing**

Here we show the image processing to generate a new map by mosaicking images whose solar incident angles are close to normal to the surface. First, we download all ISS raw images of Dione from NASA's Planetary Data System. Second, we examine all images to select images including potential rayed craters or rays. Third, the ISS raw images including candidates of potential rays were calibrated and the global map was developed by the following commands. Here we use N1649331848 as an example.

1:  ciss2isis   from=N1649331848.LBL   to= N1649331848.cub
2:  spiceinit   from=N1649331848.cub
3:  cissical   from=N1649331848.cub   to=N1649331848_cal.cub
4:  fillgap from= N1649331848_cal.cub   to=N1649331848_fill.cub
5:  photrim from=N1649331848_fill.cub to=N1649331848_trim.cub maxincidence=60
6:  cam2map from=N1649331848_trim.cub to=N1649331848_trim.map.cub

In command 3, we calibrate the ISS raw image. In command 5, we remove sections from the image, whose solar incident angle is beyond 45 degree. In command 6, we obtain the regional map whose solar incident angle is close to normal to surface (in this case, within 60 degrees). We perform the same way for other images and combine them to a mosaic, and we obtain the resultant mosaic shown in Figure 3e and Figure S2. In addition, we measure the diameter of a crater using N1649331848_fill.cub and command "qview" in ISIS3.

**Table S1.** Rayed craters on Dione

| No. | Latitude(°N) | Longitude(°W) | Diameter (km) |
|---|---|---|---|
| 1 | 49.2 | 76.3 | 36.2 |
| 2 | 31.6 | 134.7 | 9.6 |
| 3 | -7.9 | 22.8 | ~4 |
| 4 | -15.7 | 146.3 | 4.0 |
| 5 | 15.3 | 132.4 | 3.8 |
| 6 | 7.3 | 320.8 | 3.6 |
| 7 | 16.1 | 161.3 | 3.5 |
| 8 | 33.6 | 130.7 | 3.3 |
| 9 | -21.6 | 140.9 | 3.1 |
| 10 | 6.1 | 3.8 | 2.9 |
| 11 | -13.8 | 26.6 | 2.9 |
| 12 | -17.0 | 118.0 | 2.6 |
| 13 | -22.6 | 255.2 | 2.6 |
| 14 | -29.1 | 113.9 | 2.5 |
| 15 | -7.2 | 76.2 | 2.4 |
| 16 | 13.5 | 55.4 | 2.4 |
| 17 | -10.0 | 38.9 | 2.4 |
| 18 | -9.7 | 259.7 | 2.4 |
| 19 | -29.0 | 170.6 | 2.3 |
| 20 | 13.1 | 354.9 | 2.3 |
| 21 | -4.6 | 98.4 | 2.3 |
| 22 | -9.5 | 284.5 | 2.3 |
| 23 | 15.7 | 137.2 | 2.2 |
| 24 | -13.2 | 37.1 | 2.2 |
| 25 | 21.6 | 162.6 | 2.1 |
| 26 | -23.4 | 163.6 | 2.1 |
| 27 | -21.3 | 149.8 | 2.0 |
| 28 | -31.8 | 171.2 | 2.0 |
| 29 | -11.7 | 39.1 | 2.0 |
| 30 | -21.0 | 163.6 | 1.9 |
| 31 | 23.8 | 353.2 | 1.9 |
| 32 | 12.2 | 2.5 | 1.8 |
| 33 | -42.2 | 188.8 | 1.8 |
| 34 | -17.7 | 190.9 | 1.8 |
| 35 | -12.3 | 173.5 | 1.7 |
| 36 | 31.5 | 153.4 | 1.6 |
| 37 | -30.3 | 158.8 | 1.6 |
| 38 | -16.6 | 20.5 | 1.6 |
| 39 | -5.8 | 294.4 | 1.6 |

| 40 | -5.5  | 278.8 | 1.6 |
| 41 | -13.0 | 154.8 | 1.6 |
| 42 | 30.2  | 166.3 | 1.5 |
| 43 | -18.7 | 150.4 | 1.5 |
| 44 | 7.8   | 108.3 | 1.5 |
| 45 | -40.1 | 197.3 | 1.5 |
| 46 | -16.0 | 193.7 | 1.5 |
| 47 | 17.9  | 183.2 | 1.5 |

**Acknowledgement**

The authors wish to thank Dr. Michelle Kirchoff and an anonymous reviewer for their helpful comments, which significantly tightened the manuscript. This work is supported in part by Grant-in-Aid for JSPS Fellows (to NH) and JSPS KAKENHI (25120006 to HM).

**References**

Allen, C. C. (1977) Rayed craters on the Moon and Mercury. Physics of the Earth and Planetary Interiors, 15, 179-188.

Clark, R. N., J. M. Curchin, R. Jaumann, D. P. Cruikshank, R. H. Brown, T. M. Hoefen, K. Stephan, J. M. Moore, B. J. Buratti and K. H. Baines (2008). Compositional mapping of Saturn's satellite Dione with Cassini VIMS and implications of dark material in the Saturn system. Icarus, 193, 372-386.

Crater Analysis Techniques Working Group (1979). Standard Techniques for Presentation and Analysis of Crater Size-Frequency Data, Icarus, 37, 467-474.

Denk, T., G. Neukum, T. Roatsch, C. C. Porco, J. A. Burns, G. G. Galuba, N. Schmedemann, P. Helfenstein, P. C. Thomas and R. J. Wagner (2010). Iapetus: Unique surface properties and a global color dichotomy from Cassini imaging. Science, 327, 435-439.

Dones, L., C. R. Chapman, W. B. McKinnon, H. J. Melosh, M. R. Kirchoff, G. Neukum and K. J. Zahnle (2009). Icy satellites of Saturn: Impact cratering and age determination. In Saturn from Cassini-Huygens, 613-635. Springer.

Horedt, G. and G. Neukum (1984). Planetocentric versus heliocentric impacts in the Jovian and Saturnian satellite system. Journal of Geophysical Research: Solid Earth, 89, 10405-10410.

Jaumann, R., R. N. Clark, F. Nimmo, A. R. Hendrix, B. J. Buratti, T. Denk, J. M. Moore, P. M.

Schenk, S. J. Ostro and R. Srama (2009). Icy satellites: Geological evolution and surface processes. In Saturn from Cassini-Huygens, 637-681. Springer.

Kirchoff, M. R. and P. Schenk (2015). Dione's resurfacing history as determined from a global impact crater database. Icarus, 256, 78-89.

Melosh, H. J. (1989). Impact cratering: A geologic process. Research supported by NASA. New York, Oxford University Press (Oxford Monographs on Geology and Geophysics, No. 11).

Melosh, H. J. (2011). Planetary surface processes. Cambridge University Press.

Ostro, S., R. West, L. Wye, H. Zebker, M. Janssen, B. Stiles, K. Kelleher, Y. Anderson, R. Boehmer and P. Callahan (2010). New Cassini RADAR results for Saturn's icy satellites. Icarus, 206, 498-506.

Paranicas, C., E. Roussos, R. Decker, R. Johnson, A. Hendrix, P. Schenk, T. Cassidy, J. Dalton, C. Howett and P. Kollmann (2014). The lens feature on the inner saturnian satellites. Icarus, 234, 155-161.

Passey, Q. R. and E. M. Shoemaker (1982). Craters and basins on Ganymede and Callisto-Morphological indicators of crustal evolution. In Satellites of Jupiter, 379-434.

Plescia, J. (1983). The geology of Dione. Icarus, 56, 255-277.

Prockter, L. M., J. W. Head, R. T. Pappalardo, D. A. Senske, G. Neukum, R. Wagner, U. Wolf, J. O. Oberst, B. Giese and J. M. Moore (1998). Dark terrain on Ganymede: Geological mapping and interpretation of Galileo Regio at high resolution. Icarus, 135, 317-344.

Schenk, P. M., D. P. Hamilton, R. E. Johnson, W. B. McKinnon, C. Paranicas, J. Schmidt and M. R. Showalter (2011). Plasma, plumes and rings: Saturn system dynamics as recorded in global color patterns on its midsize icy satellites. Icarus, 211, 740-757.

Schenk, P. M. and S. W. Murphy (2011). The Rayed Craters of Saturn's Icy Satellites (Including Iapetus): Current Impactor Populations and Origins. In 42nd Lunar and Planetary Science Conference, Abstract #2098. Houston: Lunar and Planetary Institute.

Scipioni, F., F. Tosi, K. Stephan, G. Filacchione, M. Ciarniello, F. Capaccioni, P. Cerroni and VIMS Team (2013). Spectroscopic classification of icy satellites of Saturn I: Identification of terrain units on Dione. Icarus, 226, 1331-1349.

Shoemaker , E. M. (1962). Interpretation of Lunar Craters. Physics and Astronomy of the Moon, 283-359.

Spencer, J. R. and T. Denk (2010). Formation of Iapetus' extreme albedo dichotomy by exogenically triggered thermal ice migration. Science, 327, 432-435.

Squyres, S. W. (1980). Surface temperatures and retention of $H_2O$ frost on Ganymede and Callisto. Icarus, 44, 502-510.

Stephan, K., R. Jaumann, R. Wagner, R. N. Clark, D. P. Cruikshank, C. A. Hibbitts, T. Roatsch, H. Hoffmann, R. H. Brown and G. Filiacchione (2010). Dione's spectral and geological properties. Icarus, 206, 631-652.

Verbiscer, A., R. French, M. Showalter and P. Helfenstein (2007). Enceladus: Cosmic graffiti artist caught in the act. Science, 315, 815-815.

Wagner, R., G. Neukum, B. Giese, T. Roatsch, U. Wolf, T. Denk and Cassini ISS Team (2006). Geology, Ages and Topography of Saturn's Satellite Dione Observed by the Cassini ISS Camera. In 37th Lunar and Planetary Science Conference, Abstract #1805. Houston: Lunar and Planetary Institute.

Wagner, R. J., G. Neukum, B. Giese, T. Roatsch, T. Denk, U. Wolf and C. C. Porco (2008). Geology of Saturn's satellite Rhea on the basis of the high-resolution images from the targeted flyby 049 on Aug. 30, 2007. In 38th Lunar and Planetary Science Conference, Abstract #1930. Houston: Lunar and Planetary Institute.

Wagner, R. J., G. Neukum, U. Wolf, N. Schmedemann, T. Denk, K. Stephan, T. Roatsch and C. C. Porco (2011). Bright Ray Craters on Rhea and Dione. In 42nd Lunar and Planetary Science Conference, Abstract #2249. Houston: Lunar and Planetary Institute.

Zahnle, K., P. Schenk, H. Levison and L. Dones (2003). Cratering rates in the outer Solar System. Icarus, 163, 263-289.

Zahnle, K., P. Schenk, S. Sobieszczyk, L. Dones and H. F. Levison (2001). Differential cratering of synchronously rotating satellites by ecliptic comets. Icarus, 153, 111-129.